\documentclass[aps,pra,a4paper,twocolumn,showpacs]{revtex4}
\usepackage[dvips]{epsfig}
\usepackage[usenames]{color}
\usepackage{pstricks}
\usepackage{amsmath}
\usepackage{amssymb}
\usepackage{subfigure}
\begin{document}

\title{A Simple Quantum Model of Ultracold Polar Molecule Collisions}

\author{Zbigniew Idziaszek}
\affiliation{Faculty of Physics, University of Warsaw, 00-681 Warsaw, Poland}
\author{Goulven Qu{\'e}m{\'e}ner}
\author{John L. Bohn}
\affiliation{JILA, NIST and University of Colorado,
Boulder, CO, 80309-0440, USA}
\author{Paul S. Julienne}
\affiliation{Joint Quantum Institute, NIST and the University of Maryland, Gaithersburg, MD 20899-8423, USA}

\date{\today}

\begin{abstract}
We present a unified formalism for describing chemical reaction
rates of trapped, ultracold molecules.  This formalism reduces
the scattering to its essential features, namely, a propagation of the
reactant molecules through a gauntlet of long-range forces before
they ultimately encounter one another, followed by a probability
for the reaction to occur once they do.  In this way, the electric-field
dependence should be readily parametrized in terms of a pair of
fitting parameters (along with a $C_6$ coefficient)  for each 
asymptotic value of partial wave quantum numbers $|L,M \rangle$.  
From this, the electric field dependence of the collision rates follows
automatically.
We present examples for reactive species such as KRb, and non-reactive
species, such as RbCs.
\end{abstract}


\maketitle

\font\smallfont=cmr7

Ultracold molecules present researchers with unique physical systems 
that are a curious mixture of small and large energies, and of tiny and
enormous length scales.  Thanks to recent experimental advances, certain
molecules can be prepared in specific hyperfine states, even though they are
separated in energy only by $\sim 10^{-9}$ eV \cite{Ni08_Science,Ospelkaus10_PRL,
Danzl09_preprint}.  
Yet, upon colliding, 
these same molecules explore the $\sim 1$ eV energies afforded by their
electronic potential energy surfaces.  Similarly, the translational kinetic 
energy of these molecules, set by their temperature $T$, can be as small as 
$k_BT \sim 10^{-11}$ eV.  At these energies the force the molecules
exert on one another
can be significant on length scales that are orders of magnitude larger
than the molecules themselves. Because long-range dipolar forces
are experimentally controllable, there
has been much discussion about the prospect of either limiting or
enhancing chemical reaction rates by the simple artifice of changing
an electric field \cite{Krems_chapter}.  Indeed, an effect 
of this kind has been dramatically
demonstrated and explained recently \cite{Ni10_Nature}.
Collisions of ultracold molecules are thus, in principle, remarkably
complicated systems to understand in detail.  On the one hand, every
degree of freedom is  involved, from the hyperfine states in
which the molecules are prepared 
to the complete re-arrangement of molecules in a chemical reaction.
On the other hand, the actual number of observables may be rather small,
consisting perhaps of a handful of loss rate coefficients.
The long path connecting complex molecular Hamiltonians to what is
actually seen in the lab may indeed prove intractable from 
{\it ab initio} theory.

For this reason, it is worthwhile to find simple formulations of
collision theory at ultralow temperatures,  especially formulations
that naturally take advantage of the vast differences in energy and 
length scales present \cite{Burke98_PRL,Gao05_PRA}.  
Recently, two complementary approaches have accounted fairly well
for experiments that have observed ultracold chemical reactions of
fermionic KRb molecules.  In the first, a multichannel quantum defect 
theory (MQDT) approach
has successfully replaced the short-range physics by suitably
parametrized boundary conditions that acknowledge both a scattering
phase shift and the probability of chemical reaction \cite{Idziaszek10_PRL}.
The boundary conditions were matched to highly accurate solutions of the
long-range scattering, which in fact were carefully characterized analytically,
allowing for simple analytic formulas for scattering observables  
\cite{Gao05_PRA}.   The
second, ``quantum threshold'' (QT) approach focused on the fact that 
the molecules had to tunnel through
a centrifugal barrier with a given probability, which varied with energy in
accord with the Wigner threshold laws \cite{Quemener10_PRA}.
  By floating the value of the
tunneling probability at the barrier's peak, this method was able to 
describe in an analytic way the chemical reaction probability even in the
presence of an electric field that polarized the molecules \cite{Ni10_Nature}.

In this Rapid Communication we merge the ideas behind these approaches to arrive
at a consistent theory of ultracold polar molecule collisions.  We will 
exploit the short-range parametrization already afforded by the zero-field
MQDT approach, complemented by a numerical treatment of long-range wave function
propagation.  One main result is the classification of molecules according
to whether their scattering is universal, with loss rates that depend only
on purely long-range features of the potential energy surfaces; or else
non-universal, containing resonances that carry more detailed information
about specifics of the interactions.  These kinds of field-dependent
resonances were reported previously 
\cite{Deb01_PRA,Ticknor05_PRA,Bohn05_proceedings,Ticknor07_PRA,
Roudnev09_PRA,Roudnev09_JPB,Kotochigova10_preprint}.
 We find that the locations and contrast of these features are 
specified once the MQDT parameters 
of the short-range physics are given.  Therefore, a whole swath of
the electric-field-dependent collisional spectrum may be
simply characterized.

We begin with the Hamiltonian for interaction of two dipolar molecules in well-defined
single internal states, with reduced mass $\mu$ and intermolecular separation $R$:
\begin{eqnarray}
\label{Hamiltonian}
H = T_r + V_{\rm sr} + V_{\rm cent} + V_{\rm vdW} + V_{\rm dd}.
\end{eqnarray}
Here, $T_r$ is the radial kinetic energy; 
$V_{\rm cent} = \hbar^2 L(L+1)/2 \mu R^2$
is  the centrifugal energy corresponding to partial wave $L$;
$V_{\rm vdW} = -C_6/R^6$ is the van der Waals interaction between two
molecules, here assumed isotropic; and $V_{\rm dd} = C_3(L,L^{\prime};M_L)/R^3$ 
is the dipole-dipole interaction between the molecules, which couples
different partial waves,but which preserves the projection $M_L$
of this angular momentum onto the field axis \cite{Quemener10_PRA}.  
These three terms identify the long-range 
physics, denoted collectively as $V_{\rm lr}$.  
In addition, $V_{\rm sr}$ incorporates all short-range physics, such
as elastic and inelastic scattering, possible resonances to ro-vibrationally 
excited molecular
states, or even chemical reaction.  We will not deal explicitly with $V_{\rm sr}$
in what follows.

For the present, we are interested in collisions that may result in chemical 
reactions, rather than hyperfine-changing collisions.  We therefore ignore
all other hyperfine states besides the incident ones.  (The theory can be adapted
to include these later, however.)  In our model, scattering via the 
Hamiltonian (\ref{Hamiltonian}) is then a multi-channel problem, where the
channels are defined by the partial wave quantum numbers $L$.  Higher
values of $L$ generate higher centrifugal barriers, and thus inhibit the
passage of the molecules to short range where they can react.  Therefore,
only a  handful of $L$'s are necessary to describe chemical quenching
phenomena at ultralow temperatures.  In fact, we
consider a {\it single} potential that is constructed by diagonalizing 
$V_{\rm lr}$ in the partial wave basis at each value of $R$, in the spirit
of the Born-Oppenheimer approximation, as in Ref. 
\cite{Roudnev09_PRA}. 

We therefore reduce the problem to scattering in a single potential, albeit
one from which wave function flux can leak at small values of $R$.  Within
this model, the scattering consists of three parts: 1) molecules approach
one another and transmit some fraction of their incident flux through
the long-range potential $V_{\rm lr}$, to arrive at an intermediate separation
$R_0$; 2) the molecules enter into the near zone, where they may either react
(in which case flux is lost) or else scatter back into the channel from 
which they came, generating a phase shift; 3) what's left of the molecular
flux proceeds to infinite $R$ and counts toward elastic scattering.  
In either event, scattering is defined via the diagonal scattering matrix 
element in our potential, $S$, whose magnitude may be less
that unity if a reaction has occurred.  Quite generally, elastic and quenching
scattering rate constants are given, respectively, by
\begin{eqnarray}
\label{rate_constants}
K^{\rm el}(E) = g\frac{ \pi \hbar}{\mu k}
|1-S(E)|^2 \nonumber \\
K^{\rm qu}(E) = g \frac{\pi \hbar }{\mu k}
\left(1-|S(E)|^2 \right),
\end{eqnarray}
where $k$ is the incident wave number, and $g =1,2$
according as the particles are indistinguishable or distinguishable in their 
initial channel.


The  S-matrix is characterized by a complex phase shift via
$S = e^{2i\eta}$ , which defines the
complex, energy-dependent scattering length
\begin{eqnarray}
{\tilde a}(k) = {\tilde \alpha}(k) - 
i {\tilde \beta}(k) 
= - \frac {\tan \eta(k)} {k}.
\end{eqnarray}
The real power of the quantum defect approach is that it  provides
{\it analytic} expressions for the complex scattering length for
zero-electric-field collisions \cite{Julienne09_FD}.  This follows
from a careful parametrization of standard wave functions in the long-range
potential, which is assumed to consist solely of van der Waals plus
centrifugal potentials.  As was shown in Ref. \cite{Idziaszek10_PRL}, the scattering lengths
for the lowest partial waves simplify, in the limit $k {\tilde a} \ll 1$, to
\begin{eqnarray}
\label{scattering_length_approx}
{\tilde a}_{L=0} = a + {\bar a}y 
\frac{ 1 + (1-s)^2} {i + y(1-s)} \nonumber \\
{\tilde a}_{L=1} = -2 {\bar a}_1 (k {\bar a})^2
\frac{ y + i(s-1)} {ys + i(s-2)}.
\end{eqnarray}
Here several scale parameters are used, such as the Gribakin-Flambaum
mean scattering length 
${\bar a} = 2 \pi (2 \mu C_6/\hbar^2)^{1/4}/\Gamma(1/4)^2$, and its
$p$-wave analogue 
${\bar a}_1 = {\bar a} \Gamma(1/4)^6/(144\pi^2\Gamma(3/4)^2)=1.064 {\bar a}$.
The parameters that are specific to each particular scattering problem
are instead the real part of the zero-energy scattering length, $a$,
also given in its reduced form $s = a / {\bar a}$; and the effective
short-range channel coupling strength $y$.  In the quantum-defect 
point of view, $y$ stands for the probability of chemical reaction once the
molecules get close together: when $y=0$, chemical reactions are 
 forbidden, and the scattering is purely elastic; whereas when $y=1$,
reactions occur with maximum probability \cite{Idziaszek10_PRL}.

Thus the parametrization of scattering observables follows as given
above, whereas the actual values of the parameters $s$ and $y$ will
vary from one molecule to the next, and may be determined by fitting
experimental data.  Within the theory, their values follow ultimately 
from the value and derivative of the total wave function $\psi(R_m)$ at 
a matching radius $R_m$.
In MQDT, $R_m$ represents the boundary between long and short-range, or
equivalently small and large energy scales in the relative motion
of the molecules.  Its value is conveniently chosen to be smaller than
the characteristic length ${\bar a}$, yet larger than the scale of any
significant short-range physics or chemistry.

We come now to the main point of this article.  At the matching point
$R_m$, the interaction potential is sufficiently deep that the 
potential, and more importantly the wave
function $\psi(R_m)$, are unaffected by turning the electric field on.
This reflects the physical fact that laboratory strength fields, inducing
energies on the order of $10^{-5}$ eV by polarizing the molecules,
has no effect whatsoever on the eV-scale chemical reaction processes,
or the perhaps 0.1 eV depth of the interaction potential at $R_m$.
Therefore the parameters $s$ and $y$ can be defined once, at zero
electric field, and then used at all subsequent, higher fields.  It
only remains to propagate the wave function $\psi$ from $R_m$ out to 
infinity.  This part of the process is necessarily numerical, since the
MQDT parameters for a mixed van der Waals-plus-dipole interaction are
not yet characterized, and in any event are likely to be available
only numerically. 

\begin{figure}
\includegraphics[width=2.8in,angle=0]{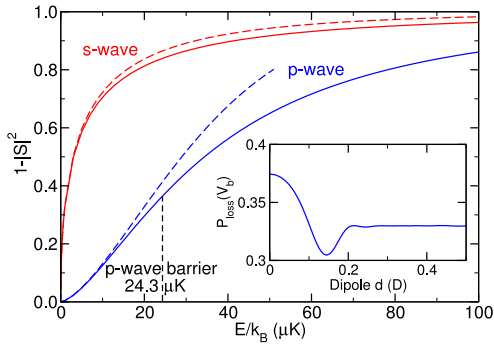}
\caption{(Color online) Probability to tunnel to short range, versus 
collision energy, for $s$- and $p$-wave collisions of KRb molecules,
assuming zero electric field and  $C_6 = 16130$ a.u. 
\cite{Kotochigova10_preprint}. The dashed lines
are the analytic, low-energy approximations $1-\exp (-4k\beta)$.
Inset: Dependence of the transmission probability, evaluated 
at the height of the centrifugal barrier, as a function of
induced dipole moment $d$.
}
\end{figure}
 
In this way the entire field-dependent scattering behavior in contemporary 
experiments can be succinctly summarized.  To set this discussion in context,
we first consider the transfer function, i.e., the probability 
that the incident flux reaches $R_m$ at all, in zero electric field.  This 
function is given by the loss probability $P_{\rm loss}=1-|S|^2$, evaluated for
unit short-range absorption probability $y=1$.  It is shown
in Fig. 1 for cold collisions of KRb molecules in zero field, 
for both $s$- and $p$-wave collisions.  
In both cases, the low-energy behavior of $P_{\rm loss}$ must follow the Wigner
threshold laws.
In the higher-energy limit, $P_{\rm loss}$ asymptotes to unity, since the 
molecules can then barrel past the comparatively weak long-range forces.
The dashed lines show approximate transfer functions that 
incorporate the lowest-order complex scattering length from 
(\ref{scattering_length_approx}), namely, $\beta_0 = \bar{a}$ for
$s$-waves and $\beta_1 = \bar{a}_1 (k \bar{a})^2$ for $p$-waves.
These approximations are adequate up to the characteristic energies
$E_0 = \hbar^2/2 \mu {\bar a}^2$ for $s$-waves, and 
$E_1 = (4 \hbar^6 / (27 \mu^3 C_6) )^{1/2}$ for $p$-waves (corresponding
to the height of the $p$-wave centrifugal barrier).  For example, for the 
KRb molecules considered in Fig. 1, $E_0 = 98 \mu$K and $E_1 = 24.3 \mu$K.

In the very low energy limit for $p$-waves, the quantum threshold model in
Ref. \cite{Quemener10_PRA} also yields the correct low-energy behavior 
of $P_{\rm loss}$,
and indeed is based on the Wigner laws.  The only remaining ingredient
within this model is to normalize $P_{\rm loss}$ to its correct value at the
height of the $p$-wave centrifugal barrier.  This value, $P_{\rm loss}(V_b)=0.37$,
is also indicated in the figure, and it is a universal value that is 
independent of the specific $C_6$ or reduced mass of the collision 
partners.  In the QT model the transmission function is
given as $P_{\rm loss} = P_{\rm loss}(V_b)(E/V_1)^{3/2}$.  Based on
this single fit parameter $P_{\rm loss}(V_b)$, the QT model is therefore
a reasonable approximation to collisions with unit absorption (or finite
absorption, by multiplying by an additional absorption probability).  
Moreover, it
is easy to evaluate at nonzero fields, by  simply evaluating $V_b$
for the correct adiabatic curve.  To a good approximation, the factor
$P_{\rm loss}(V_b)$ is only weakly-dependent
on the electric field strength, as shown in the inset to Fig. 1.  

This weak dependence can be seen, at least qualitatively, by 
replacing the actual barrier
by an artificial inverse Morse potential, constructed so as to have 
the same curvature at its barrier maximum as the actual potential
$\hbar^2 L(L+1)/ (2 \mu R^2) - C_n/R^n$.
The inverse Morse potential model can then be solved analytically as a
transmission problem~\cite{Ahmed93_PRA}, to yield 
$P_{\rm loss}(V_b) = (1-e^{- 4 \pi f})/2$, where
$f = \sqrt{L(L+1) \, 2(n-2)}/n$ .  
This approximation correctly shows that the result is independent of
the long-range coefficient $C_n$, as well as the reduced mass, but that 
it does depend 
on the partial wave $l$ as well as the character $n$ of the long-range
potential. It also shows that,
coincidentally, the transfer function at the barrier height
is the same for a van der Waals potential $n=6$ and for a dipole potential
$n=3$, for $p$-wave collisions.  Thus a weak dependence of $P_{\rm loss}(V_b)$ 
on electric field is perhaps not unexpected.  For $s$-wave collisions,
in which the long-range dipole potential scales as $C_4/R^4$
\cite{Avdeenkov02_PRA},
we would expect a stronger variation of $P_{\rm loss}(V_b)$ with electric field.

Based on these remarks, we turn now to the electric field
dependence of reactive collisions, making the assumption that
$s$ and $y$ are independent of field.
Doing so, a numerical calculation
readily produces the reaction rate constant versus the dipole moment
of the colliding species.  Examples are shown in Fig. 2 for
identical fermions (odd partial waves, Fig. 2a) and identical bosons
(even partial waves, Fig. 2b).  In both cases, the overall
tendency is for the rates to rise as the field is turned on.  This rise
is, however, more dramatic for identical fermions, which are suppressed
in zero-field by the van der Waals centrifugal barrier.

The most striking feature in these figures is the presence or
absence of resonance-like features as the dipole is increased.  
For weak short-range absorption (e.g., $y=0.1$), these features are
pronounced, and fall into regular patterns according to the angular
momentum $L$, $M$ of the dominant partial wave.  For $L>0$ these
are shape resonances behind field-dependent centrifugal barriers, 
while for $L=0$
these resonances appear as the effective long-range potential is
systematically deepened to include additional bound states 
\cite{Ticknor05_PRA,Roudnev09_PRA}. 
Vice versa, in the limit of strong
short-range absorption (e.g., $y=1$), these features are completely 
washed out.  This occurs because, at complete unit absorption,
the resonant state decays immediately; it does not survive for even
a single period of the resonance.  

The presence or absence of these resonances thus contains information on
the scattering, and in particular on the value of $y$.  
Consider, for example, KRb molecules, which are chemically reactive
at zero temperature \cite{Zuchowski10_preprint,Meyer10_preprint}, 
and should thus possess large absorption
probabilities $y$.   The points in Fig. 2a show the data from the KRb
experiment \cite{Ni10_Nature}.  
The best fit value to these data yields $y=0.83$,
consistent with large, if not perfect, absorption (red line).
Interestingly, a fit to the zero-field data alone instead yields a value
$y=0.4$ \cite{Idziaszek10_PRL}.  Thus the electric-field dependence of reaction
rates (``electric-field spectroscopy,'' \cite{Bohn05_proceedings}) 
is a valuable component in accurately determining the parameters that
govern scattering.  On the other hand, another class of molecules,
typified by RbCs, are not chemically reactive at low temperature 
\cite{Zuchowski10_preprint},
and will yield $y=0$, or very nearly so.  Here the resonances should 
show up clearly, in either elastic scattering, hyperfine-changing
collisions, or perhaps in three-body losses.  Note that the resonances
are likely to be quite narrow in this case \cite{Roudnev09_JPB}.

\begin{figure}
\includegraphics[width=3.0in]{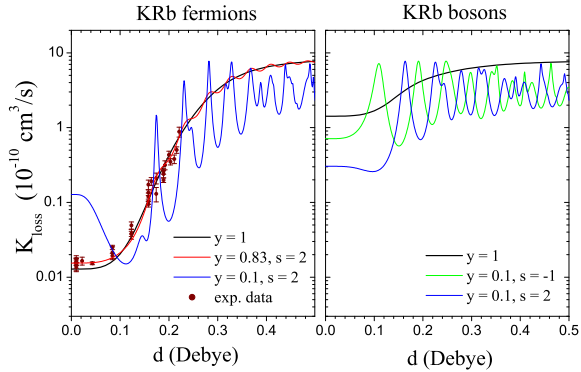}
\caption{(Color online) Dependence of chemical reaction rates $K_{\rm loss}$
on dipole moment $d$ for identical fermionic (a) or bosonic (b) KRb
molecules.  In the case of unit reaction probability, $y=1$, this variation
takes a universal form independent of details of the short-range physics.
For $y<1$, non-universal resonances appear that reveal more details of
the short-range interaction.  The data from \cite{Ni10_Nature} (points)
are well-fit by near-universal scattering, $y=0.83$
}
\end{figure}

When the resonances do appear, their positions result from phase shifts
at short range.  We illustrate this in Fig. 2b, where resonances are
shown for two alternative values of $s$.  In this figure the same
value of $s$ is used for all angular momentum channels $(L,M)$,
but in fitting real-life data this is probably not the case, and each
family of resonances specified by $(L,M)$ will likely contribute
its own complex phase shift.  Although the overall positions of the resonances
are not specified, nevertheless their relative  spacings follow
a specific pattern.  For example, for $s$-wave resonances the
location in field ${\cal E}(n)$ of the $n$-th resonance takes
the form \cite{Bohn05_proceedings}
\begin{eqnarray}
{\cal E}(n) = {\cal E}^{\prime} \sqrt{ \frac{n_0 + n}{n_{\infty}-n}},
\end{eqnarray}
where ${\cal E}^{\prime}$ is an overall scaling, $n_{\infty}$ is
related to the scattering phase shift at threshold, and 
$n_0$ is akin to a short-range quantum defect parameter.  This approximation
was derived using semicalssical arguments, which are also applicable to
the higher-partial wave case \cite{Roudnev09_JPB}.  The fully
quantum treatment described here should afford an accurate,
complete parametrization of these features.

We gratefully acknowledge support from an AFOSR MURI on Ultracold
Molecules, and a Polish Government Research Grant

\end{document}